\documentclass{article}

\usepackage{arxiv}
\usepackage[utf8]{inputenc} 
\usepackage[T1]{fontenc}    
\usepackage{hyperref}       
\usepackage{url}            
\usepackage{booktabs}       
\usepackage{amsfonts}       
\usepackage{nicefrac}       
\usepackage{microtype}      
\usepackage{graphicx}
\usepackage{natbib}
\usepackage{doi}
\usepackage{placeins}
\usepackage{amsmath} 
\usepackage{xcolor}
\usepackage{xurl}   

\title{Exploring the periodicity of flight patterns}

\author{\href{https://orcid.org/0000-0002-0243-9776}{\includegraphics[scale=0.06]{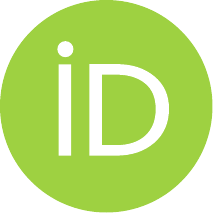}\hspace{1mm}Sarah M.~Coleman} \\
	Department of Statistics and Data Sciences\\
	The University of Texas at Austin\\
	Austin, TX 78712 \\
	\texttt{sarah.coleman@utexas.edu} \\
	\And
	\href{https://orcid.org/0000-0002-7122-1463}{\includegraphics[scale=0.06]{orcid.pdf}\hspace{1mm}H. Sherry~Zhang} \\
	Department of Statistics and Data Sciences\\
	The University of Texas at Austin\\
	Austin, TX 78712 \\
	\texttt{hsherryzhang@utexas.edu} \\
	\AND
    \href{https://orcid.org/0000-0002-1901-4301}{\includegraphics[scale=0.06]{orcid.pdf}\hspace{1mm}
	Lydia R.~Lucchesi} \\
	Department of Statistics and Data Sciences\\
	The University of Texas at Austin\\
	Austin, TX 78712 \\
	\texttt{lydia.lucchesi@austin.utexas.edu} \\
	\And
    \href{https://orcid.org/0000-0003-4183-205X}{\includegraphics[scale=0.06]{orcid.pdf}\hspace{1mm}
	Saptarshi Roy} \\
	Department of Statistics and Data Sciences\\
	The University of Texas at Austin\\
	Austin, TX 78712 \\
	\texttt{saptarshi.roy@austin.utexas.edu} \\
}

\date{}

\renewcommand{\headeright}{}
\renewcommand{\undertitle}{}
\renewcommand{\shorttitle}{Exploring the periodicity of flight patterns}

\hypersetup{
pdftitle={Exploring the periodicity of flight patterns},
pdfsubject={JSM, flight},
pdfauthor={Sarah M.~Coleman, H. Sherry~Zhang, Lydia R.~Lucchesi, Saptarshi Roy},
pdfkeywords={airline flight patterns},
}

\begin{document}

\maketitle

\begin{figure*}[ht]
    \centering
    \includegraphics[width =0.99\linewidth]{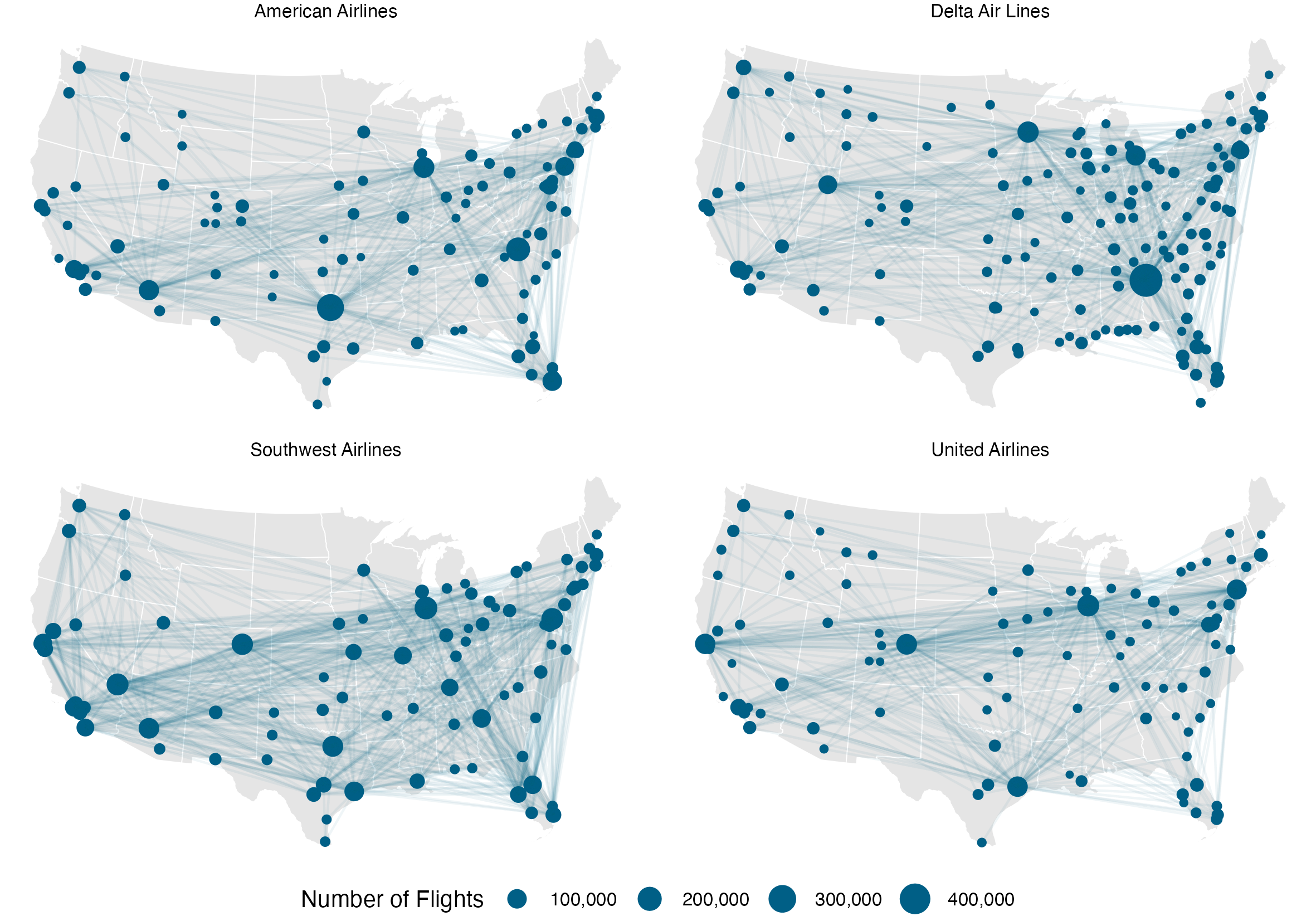}
    \caption{Flight networks for the ``big four'' U.S. airlines in 2017: American Airlines, Delta Air Lines, Southwest Airlines, and United Airlines. Points represent airports, with size proportional to the total number of flights operated by the airline at that airport during the year. Lines represent routes offered by each airline between airports.}
    \label{fig:flight-map}
\end{figure*}

Each year the American Statistical Association (ASA) hosts the Annual Data Challenge Expo, which tasks participants with analyzing a given dataset and presenting their work at the Joint Statistical Meeting (JSM). The 2025 Data Challenge Expo tasked participants with analyzing over 35 years of commercial flight data from the United States Bureau of Transportation Statistics (BTS). These data provide extensive geographic coverage and operational details for the U.S. domestic aviation market. For millions of past flights, there is information about the flight’s date, origin, destination, carrier, plane, departure, and arrival. Figure~\ref{fig:flight-map} visualizes domestic routes for four major carriers in the year 2017, to help illustrate the volume of the available flight data. 

In this article, we present our analysis for the 2025 JSM Data Challenge Expo. We chose to explore patterns in the daily scheduling of departures and arrivals across airlines, airports, and time. In doing so, we observed distinct scheduling ``waves,'' or periodic structures at major airline hubs as well as large Federal Aviation Administration (FAA) hubs. In the remainder of this article, we detail the process of visualizing periodicity in flight scheduling as well as quantifying it through the calculation of Shannon entropy. An additional element to the 2025 Data Challenge Expo is the incorporation of a second dataset, to be decided by the participants. We detail the use of a BTS dataset with passenger enplanement (boarding) information to determine Federal Aviation Administration (FAA) hub classification (as opposed to airline-specific hubs). Furthermore, we discuss results from this visual and quantitative analysis, highlighting noticeable differences in the scheduling periodicity and entropy across airports, for the ``big four'' or four largest carriers, in U.S. aviation: American Airlines, Delta Air Lines, United Airlines, and Southwest Airlines.

\section*{Exploring The Data}

With so much data to consider, we decided to start small and focus on one data year, airport, and airline at a time. Namely, for a given airport and airline, when do flights generally take off and land over the course of a day? 

\begin{figure*}[ht]
    \centering
    \includegraphics[width=0.99\linewidth]{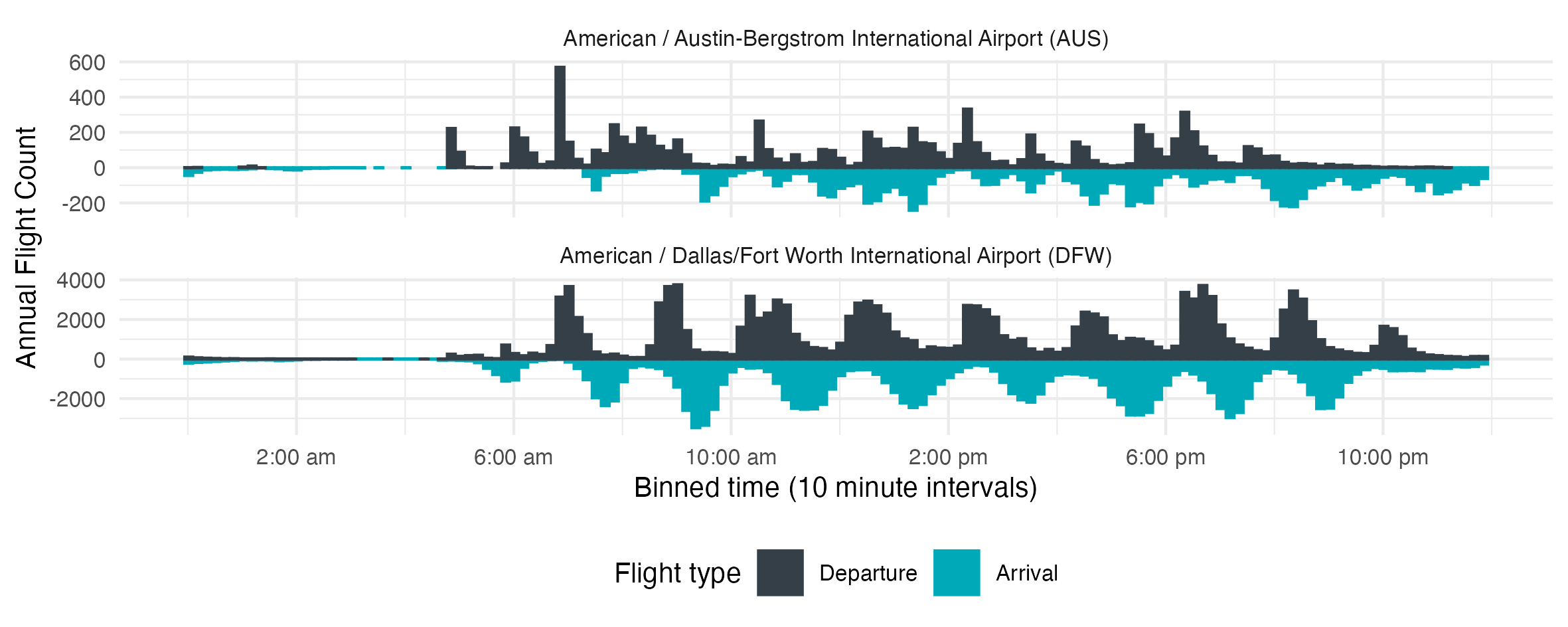}
    \caption{Arrival and departure counts for flights landing at and departing from Austin-Bergstrom International Airport (AUS) and Dallas Fort Worth International Airport (DFW) in 2017. Arrival/departure times are aggregated into 10-minute intervals and arrival counts are shown as negative values for display purposes. AUS exhibits higher departure volumes in the early morning and higher arrival volumes late at night, while DFW shows an alternating cyclic pattern in arrivals and departures at a frequency of approximately two hours.} 
    \label{fig:AUS-DFW}
\end{figure*}

Our team is located in Austin, Texas, so we first investigated Austin-Bergstrom International Airport (AUS). We also were interested in Dallas Fort Worth International Airport (DFW), the largest airport in Texas, as a reference airport for comparison. For a data year, we selected 2017 because it is relatively recent however unaffected by events such as the 2020 COVID-19 pandemic. For our initial airline selection, we chose American Airlines (AA), which has a well-known presence in Texas given its DFW hub. Here, the term ``hub'' refers to an airport that is publicly defined, by a specific airline, as one of its primary operating bases. Airlines often direct a large number of their multi-segment flights through their own hubs, with these intermediate stops known as layovers. To visualize the number of American Airlines flights in 2017, for both the AUS and DFW airports, we generated a histogram of arrival and departure times binned by 10-minute intervals. These histograms are shown in Figure~\ref{fig:AUS-DFW}. We selected 10-minute intervals for this figure, as well as of the rest of the analyses and visualizations, as this seemed to be a reasonable resolution given the data.

Beyond the much higher flight count at DFW relative to AUS, the two airports in Figure~\ref{fig:AUS-DFW} show quite interesting temporal patterns throughout the day. Specifically, AUS has a higher volume of departures in the early morning before 8:00 a.m. and arrivals in the late evening after 9:00 p.m. On the other hand, DFW seems to display an alternating arrival and departure cycle occurring approximately every two hours. This cycle was unexpected to us and became the main focus of the rest of our analysis. Importantly, we confirmed that this cyclic or wave-like pattern for AA flights at DFW is robust and observable across various sub-half-hour temporal aggregations beyond 10 minutes. 

\section*{Visualizing Flight Periodicity}

\begin{figure*}[!ht]
    \centering
    \includegraphics[width=0.99\linewidth, height=14cm]{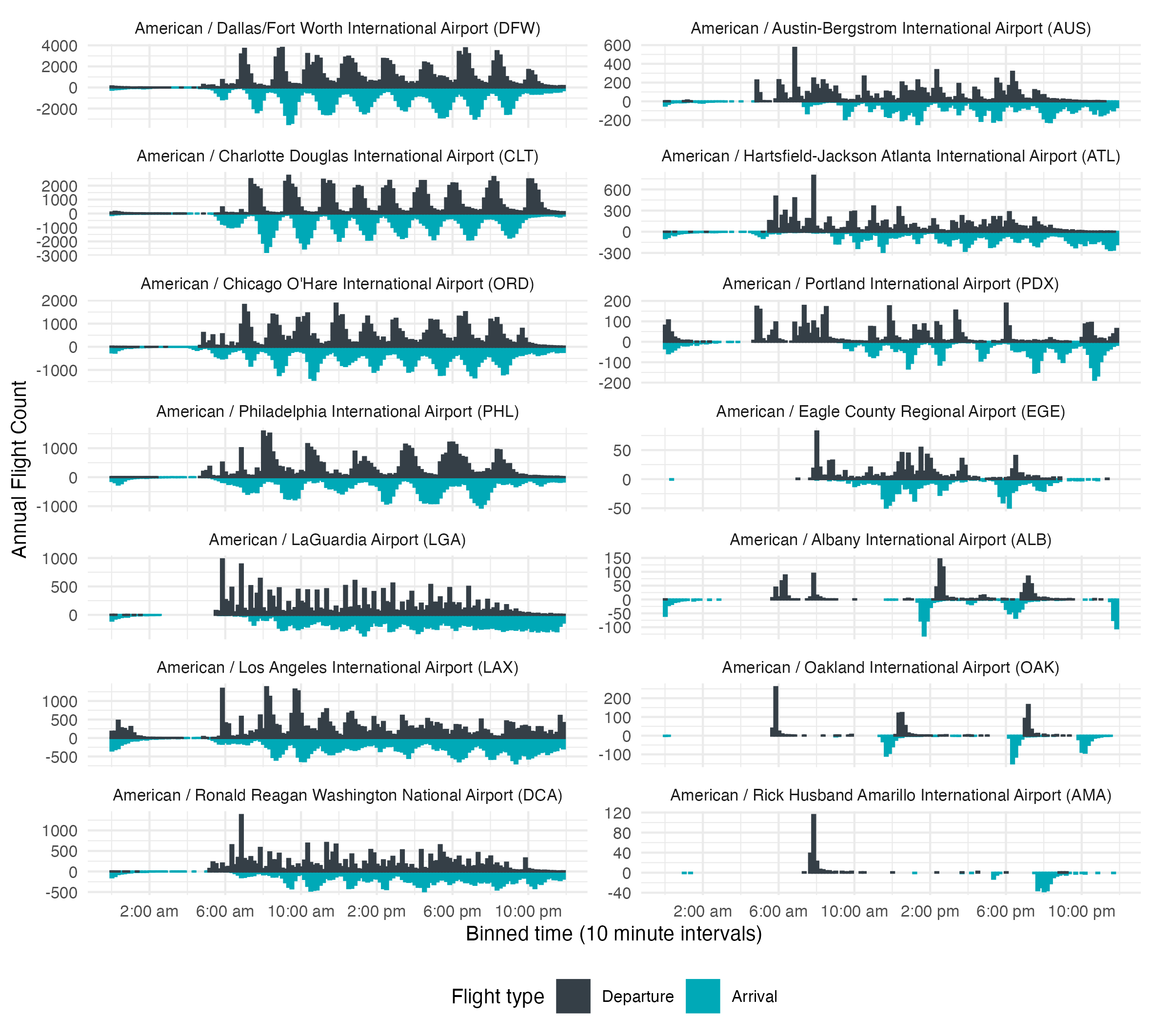}
    \caption{A similar visualization as shown in Figure~\ref{fig:AUS-DFW}, extended to a broader set of airports served by American Airlines (AA) in 2017. Each airport in the left column of this figure is an American Airlines hub. Some American Airlines hubs (DFW, CLT, ORD, and PHL) show clear alternating periodic arrival and departure patterns at a frequency of approximately two hours. Other American Airlines hubs, such as LGA, LAX, and DCA, appear to have a more frequent and/or less periodic arrival and departure pattern. Similar to Austin (AUS), other non-American hub airports (ATL, PDX, and EGE) show patterns with peaked departures in the early morning and arrivals late at night. Interestingly, as West Coast airports, PDX and LAX show a small peak in night flights. This may allow eastbound flights to land before overnight curfews at destination airports. For smaller airports (ALB, OAK, and AMA), activities are less continuous, but clustered arrival and departure patterns are observed when flights occur.} 
    \label{fig:aa-more}
\end{figure*}

The wave-like pattern in both the AUS and DFW airports caught our attention, raising several questions to consider next. Is this pattern specific to airlines such as American Airlines or airports such as DFW? Is the size of an airport related to the presence of this wave-like pattern? Is the presence of this pattern related to DFW's status as a major American Airlines hub? To investigate, we extended the visualization in Figure~\ref{fig:AUS-DFW} to include a broader set of airports that vary in flight volume. We made sure to include additional American Airlines hub airports beyond DFW, as well as airports that do not serve as American Airlines hubs. Figure~\ref{fig:aa-more} displays this information, visualizing the departures and arrivals at fourteen airports. The visualization includes major American Airlines-specific hubs (\textit{e.g.}, Charlotte Douglas International Airport, or CLT). It also includes airports with high flight volumes that serve as hubs for multiple airlines (\textit{e.g.}, Los Angeles International Airport, or LAX), as well as smaller regional airports (\textit{e.g.}, Albany International Airport, or ALB).

In Figure~\ref{fig:aa-more}, the cyclic pattern observed at DFW is also seen in some other American Airlines hubs, including Charlotte Douglas International Airport (CLT), Chicago O'Hare International Airport (ORD), and Philadelphia International Airport (PHL). In contrast, three other AA hubs---LaGuardia Airport (LGA), Los Angeles International Airport (LAX), and Ronald Reagan Washington National Airport (DCA)---which are also hubs for other airlines, show more continuous arrival patterns and more frequent departure activities throughout the day. The peaked early-morning departures and late-night arrivals observed at AUS also appear at other non-hub airports such as Hartsfield-Jackson Atlanta International Airport (ATL), Portland International Airport (PDX), and partially in Eagle County Regional Airport (EGE). Airports with lower annual flight counts, including Albany International Airport (ALB), Oakland International Airport (OAK), and Rick Husband Amarillo International Airport (AMA), show less continuous activity, but still exhibit morning and late-night peaks, with arrivals and departures occurring in clusters. The West Coast airports LAX and PDX show small peaks in late-night or overnight departures. 

It is clear that there is some relationship between American Airlines hubs and the periodic pattern of flight arrivals and departures, however the association is not entirely clear. To continue our exploration, we were next curious if this strong cyclic pattern is also observed for other ``big four'' airlines at their major hubs. To determine what constituted a major hub for an airline, we used the information we could find on each airline's website. Delta Air Lines (DL) flights exhibit the same pattern at major DL hubs like Hartsfield-Jackson Atlanta International Airport (ATL) and Minneapolis-Saint Paul International Airport (MSP). The pattern also holds for United Airlines (UA) flights at major UA hubs, including Chicago O'Hare International Airport (ORD), Denver International Airport (DEN), and George Bush Intercontinental Airport (IAH). 

Research into Southwest Airlines hubs lead us to an interesting discovery: they don't exist, at least not in the same way as they do at United Airlines, Delta Air Lines and American Airlines. To explain the difference, we must provide a bit of background about airline flight strategy. To handle large volumes of passengers, many airlines, such as American, Delta and United, have adopted a ``hub-and-spoke'' transportation network. In this strategy, departing flights are routed through major (hub) airports via connecting layover flights before arrival at the final (spoke) destination. However, Southwest Airlines instead uses a ``point-to-point'' approach, prioritizing flights between pairs of cities, or points, such as Denver, CO (Denver International Airport /DEN) and Albuquerque, NM (Albuquerque International Sunport/ABQ). This means that even with a large Southwest Airlines presence at a given airport, Southwest will not self-identify hubs the way the other major airline companies do. However, it would be very convenient if some sort of airport size classification did exist, which could apply to airports uniformly across all airlines. This would allow us to better identify hubs and continue exploring periodicity. This brings us to our second dataset.

\section*{Federal Aviation Administration (FAA) Hubs}

Separate from airline-specific hubs, the Federal Aviation Administration (FAA) classifies airports into different hub types based on the yearly amount of passengers who board flights at that airport, also called passenger enplanements. It's important to note that these FAA hub classifications apply to entire airports (\textit{i.e.}, they are not airline specific), unlike hubs as described previously. The FAA classifies airports into one of four hub categories based on annual passenger enplanement thresholds: large hubs, medium hubs, small hubs, and nonhubs. Large hubs handle the most traffic, with at minimum 1\% of annual passenger enplanements in the United States. In other words, at large hub airports, at least 1\% of all yearly commercial flight traffic departs from that airport. Medium hubs have between 0.25\% and 1\% of annual enplanements, and small hubs have between 0.25\% and 0.05\%. Airports with fewer than 0.05\% of annual enplanements are classified as nonhubs. Given the annual number of domestic passenger enplanements at all domestic airports, these percentage thresholds can be used to obtain the FAA hub classification of each airport.

The Bureau of Transportation Statistics provides information about the number of passenger enplanements in their ``T-100D-SEGMENT-ALL-CARRIER'' dataset. These data are unavailable in the original Data Challenge Expo dataset, which is focused more on flight scheduling information. We used this additional dataset to calculate the FAA hub classification of each airport for the year 2017 based on the enplanement thresholds described in the previous paragraph. Now, we are able to explore the association between airline hub and FAA hub airports, as well as how these hub types are associated with the periodic flight patterns we've already observed. 

We would expect a lot of overlap between airports classified as large FAA hubs and airline-specific hubs, since both are associated with increased flight traffic. It turns out that this is generally the case. An example airport is LAX (Los Angeles International Airport), which is a large FAA hub, as well as a major hub for American Airlines, Delta Air Lines, and United Airlines. However, not all large FAA hubs are hubs for multiple ``big four'' airlines. For example, Dallas Fort Worth International Airport (DFW), is a large FAA hub but only serves as a hub for American Airlines. Additionally, some airline hubs are not large FAA hubs and instead have strategic or regional importance. For example, Antonio B. Won Pat International Airport, or Guam International Airport (GUM), is not a large FAA hub although it is a United Airlines hub. 

Up to this point we have only been observing the periodicity of flight patterns visually. As we begin to dive deeper into the relationship between FAA hubs, airline hubs, and flight arrival and departure patterns, it is helpful to develop a more quantitative approach. Specifically, it would be helpful to calculate a single metric representing the periodicity of flight arrivals and departures, which could then be compared across different airports, airlines, and both FAA and airline-classified hubs. 

\section*{Calculating the entropy of periodic flight data}

``Entropy'' is a concept originating from thermodynamics describing the level of disorder or randomness of a system. It is so important to our understanding of how the world works that both the second and third laws of thermodynamics involve entropy. Broadly speaking, the more disordered a system is, the higher its entropy. Conversely, the more ordered, the lower the entropy. For example, imagine a neatly organized card deck stacked on a table compared to a card deck strewn on the floor in a chaotic fashion. Of these two example states for the deck of cards, the neatly stacked state has a lower entropy. Further, the spread out card deck has a higher entropy, or increased disorder. Over time, the concept of entropy has been applied to other fields, such as signal processing and machine learning. Here, entropy also describes randomness or disorder, but of data. Think about turning on a radio and hearing a bunch of static noise before switching to a melodic tune. The static noise has higher entropy than the melody, not because of differences in volume, but because static noise is more chaotic and random than the repeated rhythms and sounds of music.

Although both the deck of cards and radio example illustrate entropy broadly and qualitatively, entropy in practice is quantitative and field specific. For the purposes of this analysis, we are interested in the entropy of periodic signals, specifically the periodic flight arrival and departure histograms. The entropy of periodic signals was introduced by Claude Shannon in his 1948 paper ``A Mathematical Theory of Communication'' and therefore is often referred to as Shannon entropy. Calculating Shannon entropy of flight arrival and departure patterns is a three-step process. Broadly, the data are smoothed, a mathematical transformation called a Fourier transform is applied, and then finally Shannon entropy is calculated. In the first step, the discrete flight histogram data are smoothed using a technique called a spline fit. This is analogous to a woodworker sanding down rough edges to achieve a smooth and continuous surfaces. Mathematically, spline fits are a pre-processing technique which can be used to extrapolate discrete observations.

\begin{figure}[ht]
    \centering
    \includegraphics[width=0.5\linewidth]{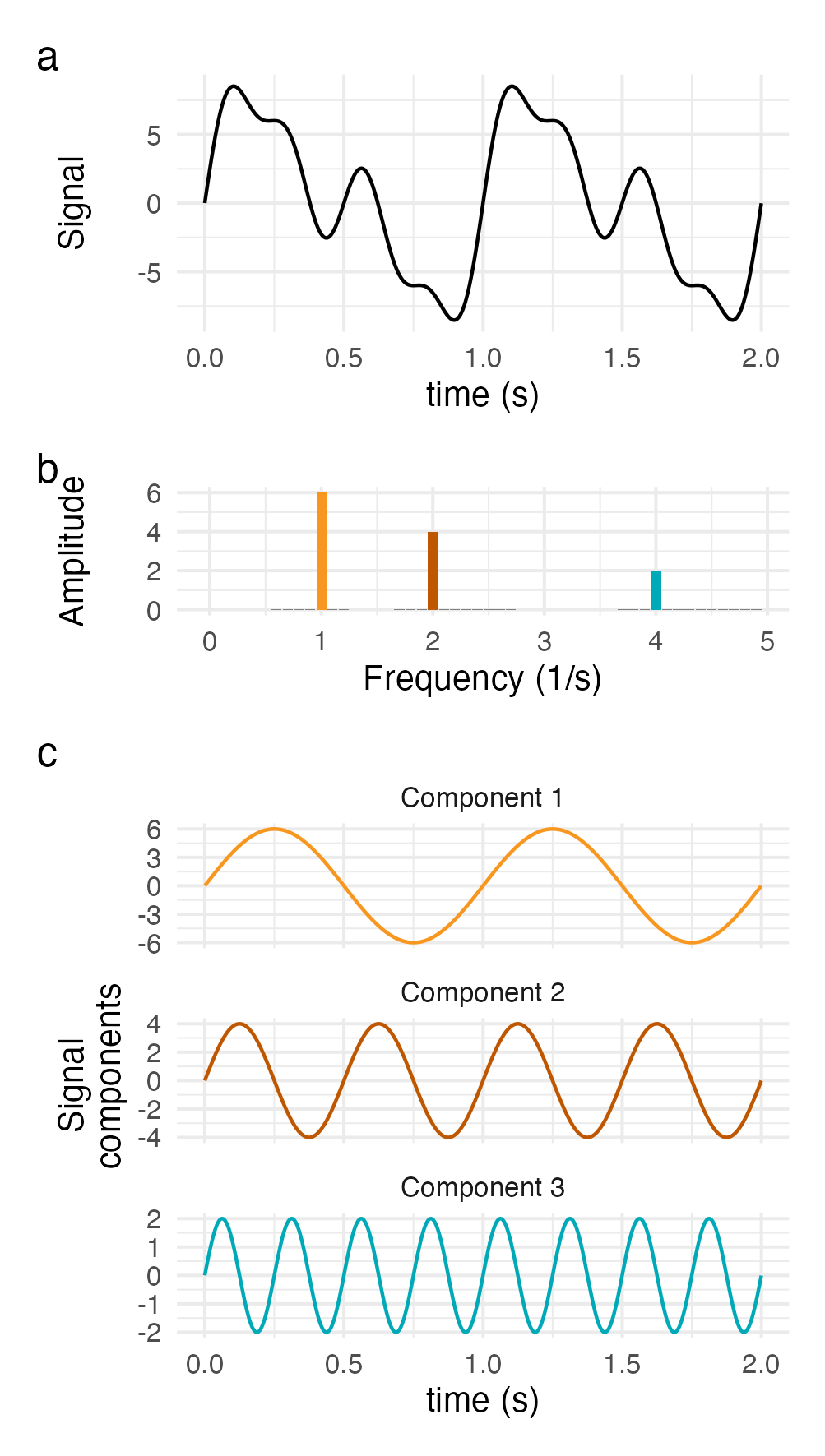}
    \caption{The Fast Discrete Fourier Transform (FFT), conceptually. An example periodic signal is shown in (a). (b) represents the FFT of (a), showing the different frequencies and amplitudes which, when combined, comprise the signal. There are three main frequency components, which are plotted in order of amplitude in (c).}
    \label{fig:fft}
\end{figure}

In the second step, a Fast Discrete Fourier transform (FFT) (Figure~\ref{fig:fft}) decomposes the smoothed signal into its different frequency components. The amount that each frequency contributes to the overall signal (mathematically, the modulus of each frequency) is related to its amplitude, or strength. This is analogous to separating an orchestral performance into the sound made by each individual instrument. In this metaphor, each frequency represents a single instrument while the amplitude represents the volume of that instrument. When the entire orchestra plays, the instruments that are the loudest have the highest amplitude, and are most easily heard by the audience. In FFT, frequencies with the largest amplitudes are called dominant frequencies and make up the majority of the signal. Interestingly, signal processing techniques like FFT are used in the identification of music through music discovery mobile phone applications like Shazam.

\begin{figure}[ht]
    \centering
    \includegraphics[width=0.5\linewidth]{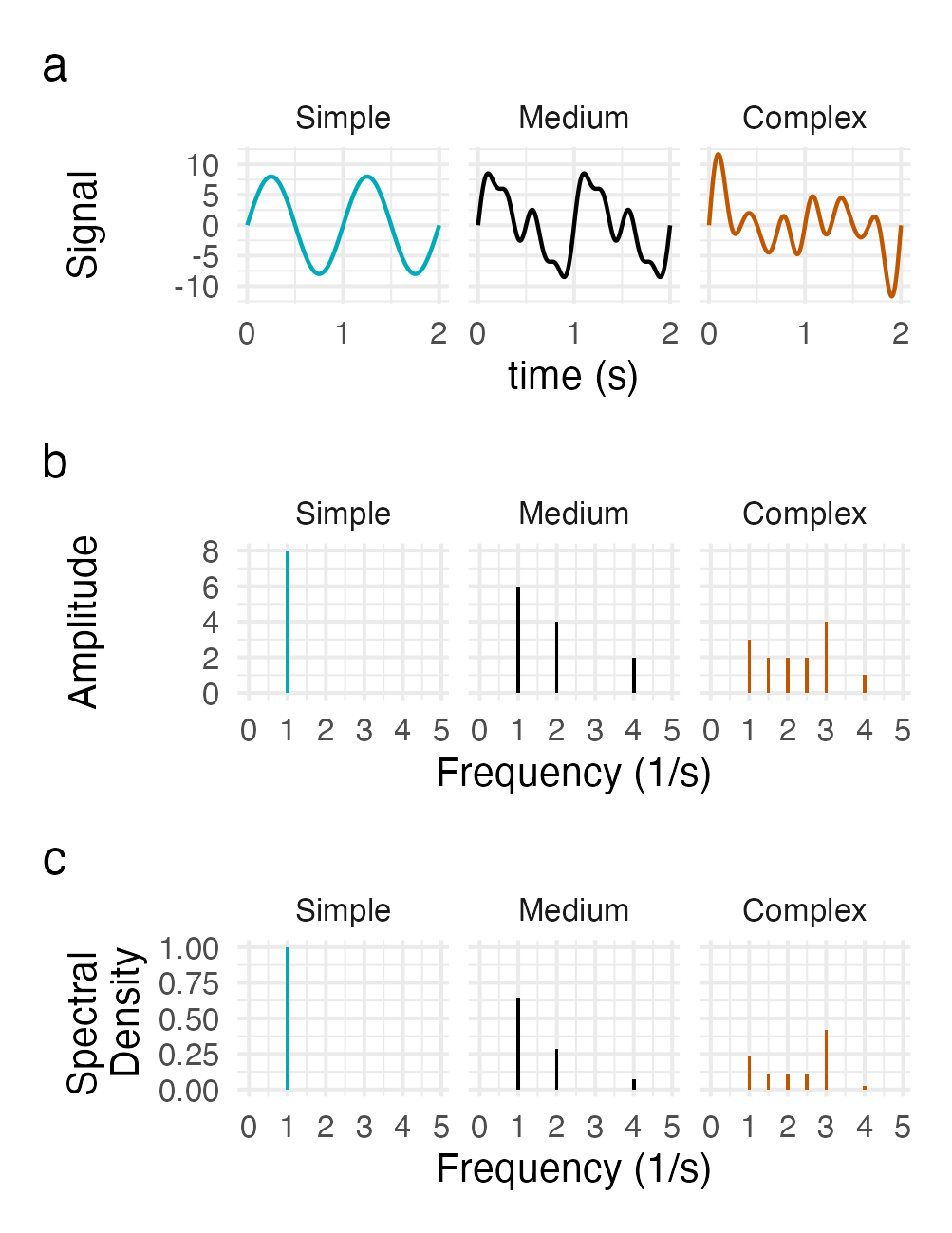}
    \caption{Conceptual visualization of the Fast Discrete Fourier transform and Power spectral density of three different signals. Each signal is shown in (a), while (b) is the amplitude and frequency of its Fourier transform, and (c) is the spectral density. The Medium signal is the same as the signal in Figure~\ref{fig:fft}. The Shannon entropy is $<0.01$ for the Simple signal (a pure sine wave), 0.83 for the Medium signal, and 1.51 for the Complex signal.}
    \label{fig:fft_to_psd}
\end{figure}

In the third and final step of the process, Shannon entropy is calculated from the spectral density of all the frequencies from the FFT (see Figure~\ref{fig:fft_to_psd}). Spectral density $p(x)$ is simply the normalized square of the amplitude $A(x)$ of each frequency, so that all $p(x)$ for the entire spectra sum to 1. Mathematically, this is written as: 

\[
    p(x) = \frac{A(x)^2}{\sum A(x)^2}
\] 

Spectral density can also be thought of as the relative distribution of each frequency, similar to a probability mass function (pmf). Shannon entropy ($H$) is equal to the sum, over all frequencies, of the negative power of each frequency $p(x)$ times its natural logarithm:

\[
    H = - \sum p(x) \log p(x)
\]

If only a few frequencies have high power, or amplitude, the Shannon entropy of the signal will be low. But if many frequencies have equal power, the Shannon entropy will be higher. Of the three example signals plotted in Figure~\ref{fig:fft_to_psd}, the relative ranking of Shannon entropy is: Simple < Medium < Complex. This is consistent with intuition, because the more complicated signal visually has a higher entropy. Connecting back to the flight departure and arrival patterns, highly periodic patterns will have low relative entropy while more chaotic or unstructured patterns will have higher entropy.

\begin{figure}[ht]
    \centering
    \includegraphics[width=0.5\linewidth]{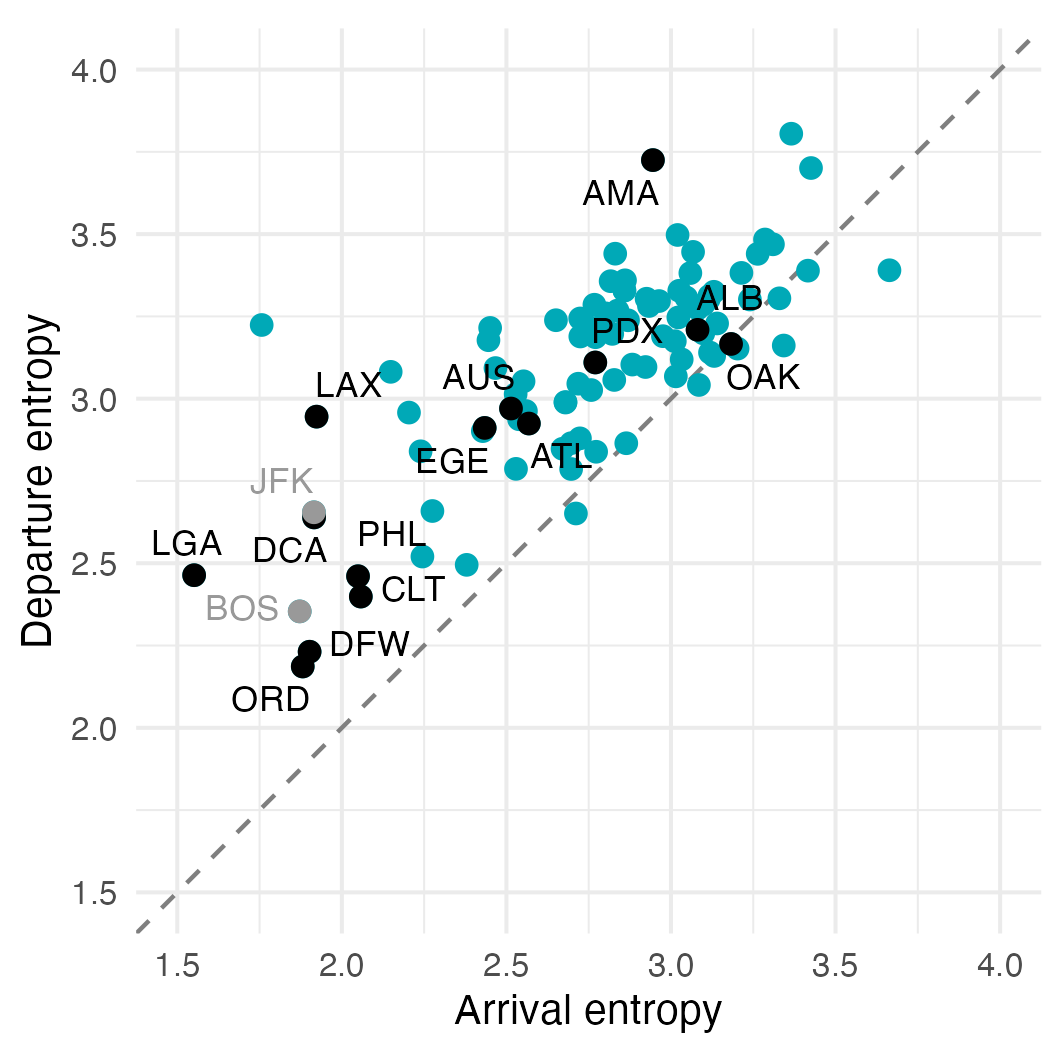}
    \caption{Arrival and departure entropy for airports served by American Airlines in 2017. Airports shown in Figure~\ref{fig:aa-more} are labeled in black, with other notable airports labeled in gray. Larger airports (DFW, CLT, ORD, PHL, LGA, LAX, and DCA) show lower entropy, while smaller airports (AMA, ALB, and OAK) tend to have higher entropy.  Airports with early-morning departure peaks and night arrival peaks (AUS, ATL, EGE, and PDX) fall in the mid-range. Airports with periodic arrival and departure patterns (DFW, CLT, ORD, and PHL) show more balanced arrival and departure entropy, whereas shared hub airports (LGA, LAX, and DCA) exhibit higher departure than arrival entropy, reflecting the departure spikes observed in Figure~\ref{fig:aa-more}. }
    \label{fig:american-entropy}
\end{figure}

\section*{Lower entropy is associated with larger airports}

Returning to the case study of American Airlines, Figure~\ref{fig:american-entropy} visualizes the arrival and departure entropy of American Airlines domestic flights at all continental U.S. airports in 2017. The airports which were plotted in Figure~\ref{fig:aa-more} are labeled in black while two other notable airports, John F. Kennedy International Airport (JFK) and Boston Logan International Airport (BOS), are shown in gray. Interestingly, the departure entropy of an airport is generally greater than its arrival entropy. Of the airports shown in Figure~\ref{fig:aa-more}, the American Airlines hubs (DFW, CLT, ORD, PHL, LGA, LAX and DCA) generally have lower arrival and departure entropy. This is consistent with our earlier observations that these airports generally have a stronger periodic pattern. Additionally, the most visually periodic American Airlines hubs (DFW, CLT, ORD and PHL) generally have lower departure entropy than the other hubs (LGA, LAX, DCA) as well as more similar arrival and departure entropy (as indicated by proximity to the dashed gray line). As expected, the non-American Airlines hub airports shown in Figure~\ref{fig:aa-more} have higher arrival and departure entropy. The smallest airports (AMA, ALB, OAK) have some of the highest overall entropies, while other mid-size airports (AUS, ATL, EGE, PDX) fall towards the middle. Satisfied with the congruence of our visual observations and entropy calculations, we next explored these results for the other ``big four'' airlines, in addition to FAA hub classifications.

\begin{figure*}[ht]
    \centering
    \includegraphics[width=\linewidth]{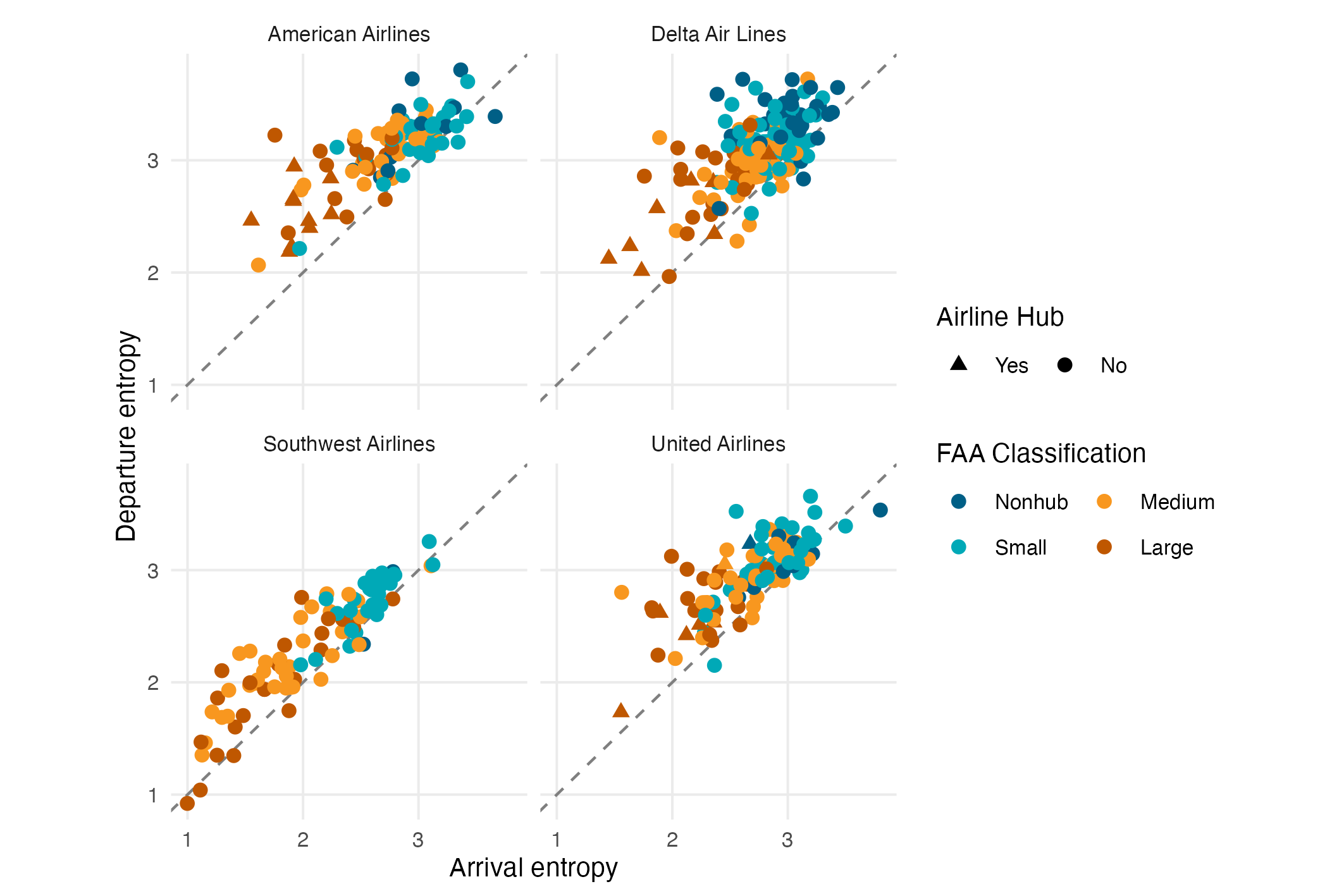}
    \caption{Extension of Figure~\protect\ref{fig:american-entropy} to each of the ``big four'' airlines in 2017. Coloring is based on FAA hub classification, where dark blue represents nonhub airports and light blue represents small hub airports. Medium hub airports are colored orange and large hub airports are colored dark orange. If an airport is a hub for a given airline, it will be represented by a triangle marker, otherwise it will have a circle marker. Because Southwest Airlines does not define hubs, none are represented. The line of equality, where arrival and departure entropies are equal, is represented by a dashed gray line. Generally, for each airline, the smallest FAA hubs (nonhub) have larger arrival and departure entropies relative to all other airports. Additionally, the largest FAA hubs (large) have some of the lowest arrival and departure entropies.}
    \label{fig:entropy}
\end{figure*}

Figure~\ref{fig:entropy} shows a scatter plot of the Shannon entropy for the arrival and departure of flights in 2017 for all airports used by the ``big four'' airlines. As in the previous figure, each point represents the entropy of the annual arrival and departure patterns of flights from a specific airport and airline. High-traffic muti-airline hub airports such as LAX will appear in each of the four panels, because all of the ``big four'' airlines operate flights which arrive and depart at LAX. In this figure, color represents the FAA hub classification, while shape represents the airline hub classification. As mentioned earlier, Southwest Airlines does not self-identify hubs, so none are shown. 

Interestingly, the FAA classification of hubs are relatively well separated for each of the four airlines, with the highest overlap appearing for United Airlines (Figure~\ref{fig:entropy}). Also, arrival and departure entropy are generally well correlated with each other, as shown by their proximity to the dashed gray line, representing y = x. Intuitively, nonhub and small airports have overall higher entropy. These airports have less overall flights, so that arrivals and departures are more likely to be chaotically spaced, because they are more dependent on the hub airports they arrive and depart at. Additionally, they are more likely to have morning or late-night peaks, as seen visually in Figure~\ref{fig:aa-more}. On the other hand, large and medium hubs generally have lower entropy. These FAA classified hubs are more likely to be airport hubs and therefore may have planned arrival and departure patterns (especially banked), explaining their lower entropy. 

Of the ``big four'', Southwest Airlines clearly has the lowest overall entropy (Figure~\ref{fig:entropy}). Additionally, the entropy of Southwest Airlines-based airports appears to be better separated by FAA classification than the other three major airlines. The result that entropy is generally lower for Southwest Airlines was unexpected and caused us to research further into flight scheduling. We suspected this entropy result may be related to differences in network strategies (Southwest uses ``point-to-point'' while others use ``hub-and-spoke''). In fact, most ``hub-and-spoke'' airlines use a ``banked-hub'' strategy to organize flights at their hubs, meaning that many flights are scheduled to arrive within a short period of time, followed by periods of few flight arrivals. This allows for quick connecting flights, consistent with ``the hub-and-spoke'' strategy. It follows that arrivals and departures at spoke airports are somewhat randomly spaced based on what is most convenient for the hub. Conversely, ``point-to-point'' airlines are more likely to use a ``rolling-hub'' strategy at airports, meaning the arrivals and departures of flights are more evenly spaced. Because the ``point-to-point'' network does not rely on layovers as intensely, this allows for airlines to distribute flights more evenly to help with other factors which may impact flight timing, such as gate availability. This even distribution of arrivals and departures for ``rolling-hub'' airports could explain why airport flight schedules at Southwest Airlines generally have lower entropy; however, further evidence is needed before a definitive conclusion can be drawn.

\begin{figure*}[ht]
    \centering
    \includegraphics[width=0.9\linewidth]{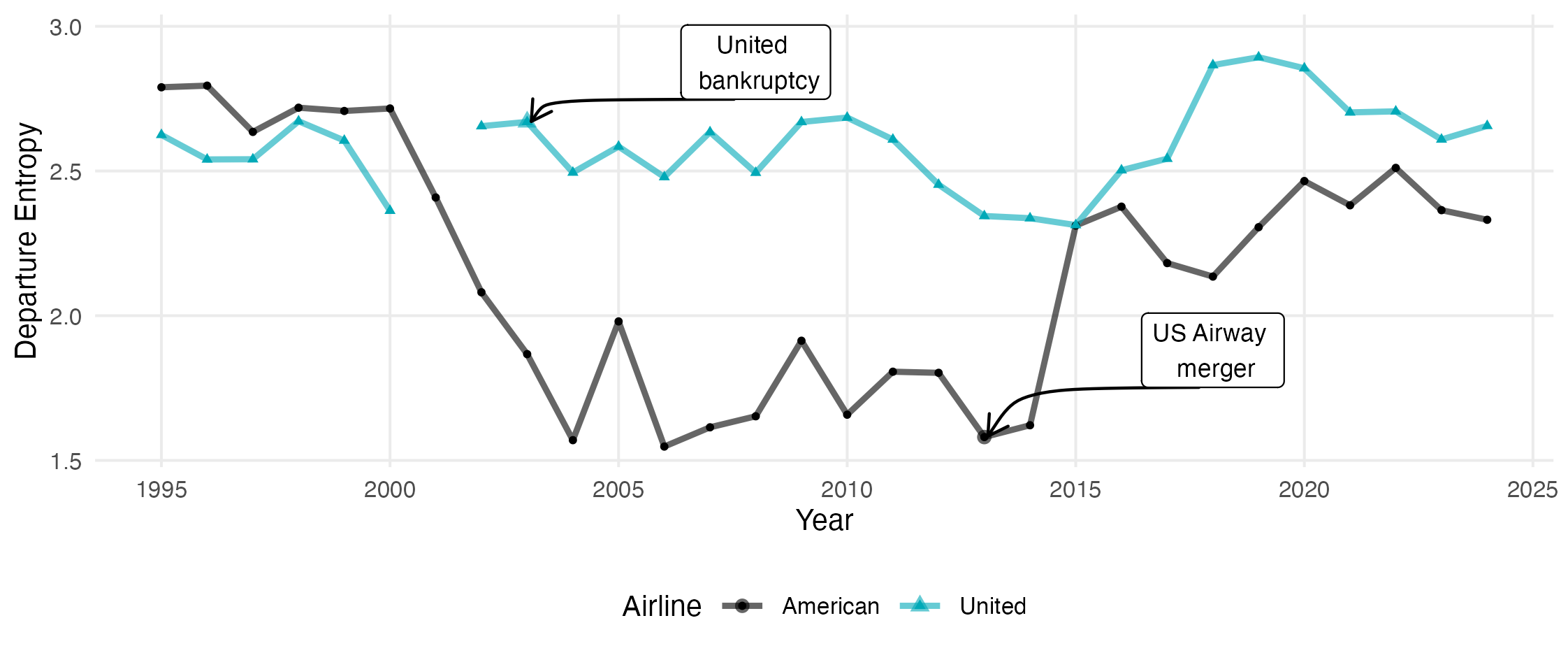}
    \caption{Case study of the change in yearly departure entropy at O'Hare International Airport (ORD) for American Airlines and United Airlines flights after recent events, which are labeled with a text box. There is no observation of departure entropy for United Airlines in 2001 due to missing data.}
    \label{fig:case_study}
\end{figure*}

\section*{Changes in entropy over time}

Up to this point, entropy visualizations have been specific to the year 2017, which was selected because it is a relatively recent year with complete data and also unaffected by large-scale flight disruptions due to the 2020 COVID-19 pandemic. However, visualizing entropy across years may be an interesting way to uncover temporal changes in airline strategy, or fights for dominance over key airport hubs. It is important to note that airlines can and do sometimes change their hub airports over long time periods for a variety of reasons. We uncovered an interesting example of changing airline strategy elucidated by entropy in the fight for dominance between United Airlines and American Airlines over O'Hare International Airport (ORD) in Chicago, Illinois. In the twentieth century, both airlines had a relatively steady presence at the airport. However, United Airlines ran into financial problems in the early 2000's, resulting in a declaration of Chapter 11 bankruptcy in 2002 (Figure~\ref{fig:case_study}). This allowed American Airlines to swoop in and reclaim ORD, as represented by a decline in departure entropy (generally associated with a larger flight count). This dominance continued relatively steadily until December 2013, when American Airlines merged with US Airways and again regrouped its strategy. This caused a restructuring of flight departures which is shown graphically in Figure~\ref{fig:case_study} by a large increase in entropy in the following year, 2014. 


\section*{Conclusion}

The scheduling of flight arrivals and departures by the each of the ``big four'' airlines in the United States---American Airlines, Delta Air Lines, United Airlines, and Southwest Airlines---is extremely complicated. Interestingly, the aggregate flight arrival and departure times at some airports over an entire year are very wave-like, or periodic, with the pattern repeating approximately every two hours. There appears to be an association with airline-specific hub status and a periodic arrival/departure structure; however, FAA-classified hubs, based on overall passenger volume, may also be related. The periodicity of the arrival and departure patterns of flights for an airline-airport pair can be quantified with a Fast Discrete Fourier Transform (FFT), and then converted into a single statistic with intuitive meaning, Shannon entropy. Associations between entropy, airport size (based on FAA hub classification) and self-declared airport hubs can then be explored within a year, or across different years. 

\section*{Additional reading} 

Bracewell, R. N. (1989). The Fourier Transform. \textit{Scientific American}, \textit{260}(6), 86–95. \url{http://www.jstor.org/stable/24987290}

Unser, M. (1999). Splines: a perfect fit for signal and image processing. \textit{IEEE Signal Processing Magazine}, \textit{16}(6), 22–38. doi:10.1109/79.799930

\end{document}